\documentclass[preprintnumbers,showkeys,amsmath,amssymb]{revtex4}
\usepackage{graphicx}
\usepackage{dcolumn}
\usepackage{bm}
\usepackage{epstopdf}
\usepackage{amsmath}
\usepackage{multirow}
\usepackage{float}
\usepackage{soul,color,xcolor}
\soulregister\upcite7
\soulregister\citep7
\soulregister\citet7
\soulregister\ref7
\soulregister\pageref7
\soulregister\and7
\soulregister\Name7

\begin{document}

\preprint{}

\title{Universal quantum multi-qubit entangling gates with auxiliary spaces}

\author{Wen-Qiang Liu$^{1,2}$, Hai-Rui Wei$^{1}$,\footnote{Corresponding author: hrwei@ustb.edu.cn} and Leong-Chuan Kwek$^{3,4,5}$}

\address{$^{1}$School of Mathematics and Physics, University of Science and Technology Beijing, Beijing  100083, China\\
$^{2}$Center for Quantum Technology Research and Key Laboratory of Advanced Optoelectronic Quantum Architecture and Measurements (MOE), School of Physics, Beijing Institute of Technology, Beijing 100081, China \\
$^{3}$Centre for Quantum Technologies, National University of Singapore, Singapore 117543, Singapore\\
$^{4}$MajuLab, CNRS-UNS-NUS-NTU International Joint Research Unit,  Singapore UMI  3654, Singapore\\
$^{5}$National Institute of Education and Institute of Advanced Studies, Nanyang Technological University, Singapore 637616, Singapore}

\begin{abstract}
Universal quantum entangling gates are a crucial building block in the large-scale quantum computation and quantum communication, and it is an important task to find simple ways to implement them. Here an effective quantum circuit for the implementation of a controlled-NOT (CNOT) gate is constructed by introducing a non-computational quantum state in the auxiliary space. Furthermore, the method is extended to the construction of a general $n$-control-qubit Toffoli gate with $(2n-1)$ qubit-qudit gates and $(2n-2)$ single-qudit gates. Based on the presented  quantum circuits, the polarization CNOT and Toffoli gates with linear optics are designed by operating on the spatial-mode degree of freedom of photons. The proposed optical schemes can be achieved with a higher success probability and no extra auxiliary photons are needed.
\end{abstract}


\keywords{\emph{quantum computation, quantum circuit, quantum gate, higher-dimensional space, linear optics}}

\maketitle

\section{Introduction}\label{sec1}


Multi-qubit quantum entangling gates have complex structures and play an important role in quantum computing \cite{universal}, quantum algorithms \cite{Gover,Shor,Long-algorithm,Grover-Toff,QFT-Toff}, quantum communication \cite{Pan-communication,QC-Long,QSDC-Sheng,QKD-Sheng,QSDC-Long,QSDC-Li,QSDC-Long1,QSDC-Long2,zhou2020measurement,QKD-Sheng1}, cryptography \cite{cryptography}, etc. \cite{book}.  In theory, multi-qubit quantum gates can be realized by sequences of two-qubit gates and single-qubit gates in a quantum circuit model  \cite{universal}.  The cost (also called complexity) of the quantum circuits usually is measured by the number of the two-qubit entangled gates involved in the quantum circuit, because they introduce more imperfections and more demands than the single-qubit gates. However, when the cost of a quantum circuit is high, it is difficult to  perform the experiments because of the low computing fidelity and limited coherence time. Moreover, the cost of a universal quantum circuit increases exponentially with the accumulation of the number of qubits. The theoretical lower bound for simulating an $n$-qubit universal quantum circuit is $(4^n-3n-1)/4$ controlled-NOT (CNOT) gates in a qubit system \cite{lower-bound}. Hence, it is crucial to find an effective method for building a universal quantum circuit in the simplest possible way.


Several matrix decomposition techniques have been introduced to optimize a large-scale quantum circuit \cite{Reck,KGD,CSD,CCD,QSD,OED,Cartan,CSD1,Sawicki1,Clements}. Two-qubit universal quantum circuits have also  been constructed with the lowest cost (resources) in qubit systems \cite{2-qubit1,2-qubit2,small-circuit,lower-bound}. However, there is still a gap between the current best result \cite{QSD} and the theoretical lower bound \cite{lower-bound} for a multi-qubit universal quantum circuit. Fortunately, Ralph \emph{et al}. \cite{T-PRA} found that the quantum circuit may be optimized further by using higher-dimensional Hilbert spaces, and this proposal was later experimentally demonstrated in optical \cite{T-NatPhy} and superconducting systems \cite{superconducting-Toffoli}.
Following this, Liu \emph{et al}. \cite{Li0,Li} reduced the cost of the $n$-qubit universal circuit to $(5/16)\times 4^n-(5/4)\times 2^n+2n$ CNOT gates when $n$ was even and $(5/16)\times 4^n-2^n+2(n-1)$ CNOTs when $n$ was odd. Liu \emph{et al}.  simplified a Fredkin gate from eight CNOTs to five CNOTs \cite{Liu1} or three qubit-qudit gates \cite{Liu2}. In addition,  higher-dimensional quantum systems have also been studied \cite{qudit-gate1,qudit-gate2} and applied in quantum computing \cite{gate-2015,OAM-gate,OAM-gate2,qudit-computing},  quantum communication \cite{d-QKD,manipulating-qutrit,multi-entanglement,entanglement-swapping,Bell,superdense-coding,d-teleportation,Teleportation}, and quantum metrology \cite{metrology}.


The Toffoli (controlled-controlled-NOT)  gate, a three-qubit conditional operation, is one of the most popular universal multi-qubit quantum gates \cite{synthesis,Sawicki2,Sawicki3}. It is also an essential component in complex quantum algorithms \cite{Gover,Shor,Long-algorithm,Grover-Toff,QFT-Toff}, quantum error correction \cite{error-correction1,error-correction2}, and quantum fault tolerance \cite{tolerance1,tolerance2}. In 1995, Barenco \emph{et al}. \cite{universal} proposed a concrete construction of a three-qubit Toffoli gate with five two-qubit entangled gates. When two-qubit gates are restricted to CNOT gates, the optimal cost of a Toffoli gate increases to six \cite{six-Toffoli}. In 2013, Yu \emph{et al}. \cite{Five1,Five2} confirmed that the minimum resource for simulating a three-qubit Toffoli gate is five two-qubit gates. In 2020, Kiktenko \emph{et al}. \cite{n-Toffoli} constructed a generalized $m$-qubit Toffoli gate with $(2m-3)$ CNOTs based on qudits. Independent of the standard decomposition-based approach,  the realization of Toffoli gate has been proposed theoretically and implemented experimentally  in superconducting circuits \cite{superconducting-Toffoli,error-correction2}, linear optics \cite{T-NatPhy,linear3,gate-2015,linear4,OAM-linear}, trapped ions \cite{ion}, atoms \cite{atom1,atom3}, and quantum dots \cite{QD}.


Ralph \emph{et al}. \cite{T-PRA,T-NatPhy} first proposed an interesting scheme for synthesizing a Toffoli gate using three qubit-qudit CNOT gates and two single-qutrit $X_A$ gates. The main idea of the works in Refs. \cite{T-PRA,T-NatPhy} was to extend temporarily the higher-dimensional subspaces on one of the controlled qubit carriers and then perform corresponding logical operations. Using the same method as Refs. \cite{T-PRA,T-NatPhy}, in this paper, we propose an alternative scheme to implement the CNOT and Toffoli gates based on the partial-swap (P-SWAP) gates  by using higher-dimensional spaces.
Specifically, $(2n-1)$ qubit-qudit and  $(2n-2)$ single-qudit gates are required to implement an $n$-control-qubit Toffoli gate. In addition, using the spatial-mode degree of freedom (DOF) of the single-photon, we design a feasible optical architecture for implementing CNOT and Toffoli gates with linear optics. Our proposals have several other advantages: (i) Our optical implementation of the CNOT gate does not require an extra entangled photon pair or a single-photon, and the success probability of the gate is enhanced. (ii) Linear optical Toffoli  gates can be constructed with a higher success probability than other existing optical schemes \cite{T-PRA,T-NatPhy,LO-Toffoli}. (iii) Our schemes are simple and  feasible with the current technology.

\section{Construction of CNOT and Toffoli gates with higher-dimensional spaces}\label{sec2}

\subsection{Synthesis of a CNOT gate using qutrits}\label{sec2.1}

\begin{figure}  [H]    
\centering
\includegraphics[width=6 cm,angle=0]{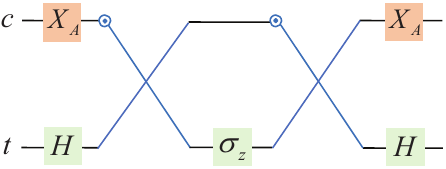}
\caption{Synthesis of a CNOT gate. The single-qutrit $X_A$ gate implements the transformation $|1\rangle \leftrightarrow |2\rangle$. The controlled node $\odot$  is turned on for the input $|0\rangle$ or $|1\rangle$. That is, a swap operation is applied to $c$ and $t$, if and only if, the control qubit $c$ is in the state $|0\rangle$ or $|1\rangle$. $H$ is a single-qubit Hardmard gate to achieve operations $|0\rangle \leftrightarrow \frac{1}{\sqrt{2}}(|0\rangle+|1\rangle)$ and $|1\rangle \leftrightarrow \frac{1}{\sqrt{2}}(|0\rangle-|1\rangle)$. $\sigma_z$ completes $\sigma_z|0\rangle=|0\rangle$ and $\sigma_z|1\rangle=-|1\rangle$.} \label{CNOT-gate}
\end{figure}

A CNOT gate with two P-SWAP gates using qutrits is shown in Figure \ref{CNOT-gate}. The gate qubits are encoded on two computational states, $|0\rangle$ and $|1\rangle$. The single-qutrit $X_A$ gate provides a three-dimensional subspace on the control qubit. In the following, we describe the construction process of our protocol in detail.

Suppose that the state of the system is initially
\begin{align}
\begin{split}             \label{eq1}
|\phi_0\rangle=\alpha_1|0_{c}\rangle|0_{t}\rangle+\alpha_2|0_{c}\rangle|1_{t}\rangle
+\alpha_3|1_{c}\rangle|0_{t}\rangle+\alpha_4|1_{c}\rangle|1_{t}\rangle.
\end{split}
\end{align}
where $\alpha_i$ ($i=1, 2, 3, 4$) are complex coefficients that satisfy the normalization condition $\sum^{4}_{i=1} |\alpha_{i}|^{2}=1$. Subscripts $c$ and $t$ denote the control and target qubits, respectively.

First, qubit $c$ undergoes a single-qutrit gate $X_A$, which introduces an ancillary state $|2\rangle$ on $c$ and completes the transformations $|1_{c}\rangle \stackrel{X_A}{\longleftrightarrow} |2_{c}\rangle$ and $|0_{c}\rangle \stackrel{X_A}{\longleftrightarrow} |0_{c}\rangle$. After the $X_A$ gate and a Hadamard ($H$) gate are applied to $c$ and $t$, the initial state $|\phi_0\rangle$ is changed to
\begin{align}
\begin{split}             \label{eq2}
|\phi_1\rangle=\frac{1}{\sqrt{2}}\big[\alpha_1|0_{c}\rangle(|0_{t}\rangle+|1_{t}\rangle)+\alpha_2|0_{c}\rangle(|0_{t}\rangle-|1_{t}\rangle)
+\alpha_3|2_{c}\rangle(|0_{t}\rangle+|1_{t}\rangle)+\alpha_4|2_{c}\rangle(|0_{t}\rangle-|1_{t}\rangle)\big].
\end{split}
\end{align}

Second, a P-SWAP gate is applied to $c$ and $t$, and it transforms $|\phi_1\rangle$ into
\begin{align}
\begin{split}             \label{eq3}
|\phi_2\rangle=&\frac{1}{\sqrt{2}}\big[\alpha_1(|0_{c}\rangle+|1_{c}\rangle)|0_{t}\rangle +\alpha_2(|0_{c}\rangle-|1_{c}\rangle)|0_{t}\rangle
+\alpha_3|2_{c}\rangle(|0_{t}\rangle+|1_{t}\rangle) +\alpha_4|2_{c}\rangle(|0_{t}\rangle-|1_{t}\rangle)\big].
\end{split}
\end{align}
Here, the P-SWAP gate performs a swap operation only between two computational states $|0\rangle$ and $|1\rangle$, that is,
\begin{eqnarray}              \label{eq4}
\begin{split}
&|00\rangle \xrightarrow{\text{P-SWAP}} |00\rangle, \qquad\;\; &|01\rangle \xrightarrow{\text{P-SWAP}} |10\rangle, \\
&|10\rangle \xrightarrow{\text{P-SWAP}} |01\rangle, \qquad\;\; &|11\rangle \xrightarrow{\text{P-SWAP}} |11\rangle, \\
&|20\rangle \xrightarrow{\text{P-SWAP}} |20\rangle, \qquad\;\; &|21\rangle \xrightarrow{\text{P-SWAP}} |21\rangle.
\end{split}
\end{eqnarray}

Third, a $\sigma_z$ operation acts on $t$ to change $|\phi_2\rangle$ to
\begin{align}
\begin{split}             \label{eq5}
|\phi_3\rangle=&\frac{1}{\sqrt{2}}\big[\alpha_1(|0_{c}\rangle+|1_{c}\rangle)|0_{t}\rangle+\alpha_2(|0_{c}\rangle-|1_{c}\rangle)|0_{t}\rangle +\alpha_3|2_{c}\rangle(|0_{t}\rangle-|1_{t}\rangle)+\alpha_4|2_{c}\rangle(|0_{t}\rangle+|1_{t}\rangle)\big].
\end{split}
\end{align}

Finally, after the P-SWAP gate, the $X_A$ gate and $H$ operation are applied to $c$ and $t$ again,  $|\phi_3\rangle$ is changed to
\begin{align}
\begin{split}             \label{eq6}
|\phi_4\rangle=&\alpha_1|0_{c}\rangle|0_{t}\rangle+\alpha_2|0_{c}\rangle|1_{t}\rangle+\alpha_3|1_{c}\rangle|1_{t}\rangle+\alpha_4|1_{c}\rangle|0_{t}\rangle.
\end{split}
\end{align}

From Equation  (\ref{eq1}) to Equation (\ref{eq6}), one can see that a CNOT gate is completed by Figure \ref{CNOT-gate},  and such a construction can be achieved in linear optics with a high success probability and without additional photons (see section \ref{Sec3}).

\subsection{Construction of the Toffoli gates with higher-dimensional spaces} \label{Sec2.2}

\subsubsection{Synthesis of three-qubit Toffoli gate using qutrits} \label{Sec2.2.1}

\begin{figure}  [H]    
\centering
\includegraphics[width=8.5 cm,angle=0]{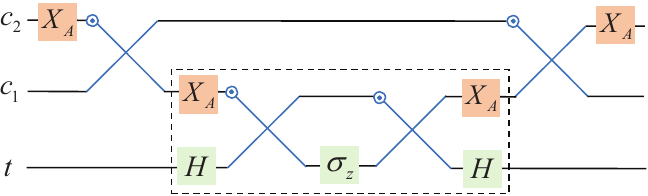}
\caption{Simplified synthesis of a three-qubit Toffoli gate with two P-SWAP and one CNOT gates. As shown in Fig. \ref{CNOT-gate}, the operations in the dotted rectangle are a CNOT gate.} \label{3-qubit-Toffoli}
\end{figure}

Based on the CNOT and P-SWAP gates, the process for implementing a three-qubit Toffoli gate with four-dimensional space  is presented in Figure \ref{3-qubit-Toffoli}.

Considering an arbitrary normalization three-qubit initial state
\begin{align}
\begin{split}             \label{eq7}
|\psi_0\rangle=&\alpha_1|0_{c_1}\rangle|0_{c_2}\rangle|0_{t}\rangle+\alpha_2|0_{c_1}\rangle|0_{c_2}\rangle|1_{t}\rangle
               +\alpha_3|0_{c_1}\rangle|1_{c_2}\rangle|0_{t}\rangle+\alpha_4|0_{c_1}\rangle|1_{c_2}\rangle|1_{t}\rangle \\
               &+\alpha_5|1_{c_1}\rangle|0_{c_2}\rangle|0_{t}\rangle+\alpha_6|1_{c_1}\rangle|0_{c_2}\rangle|1_{t}\rangle
               +\alpha_7|1_{c_1}\rangle|1_{c_2}\rangle|0_{t}\rangle+\alpha_8|1_{c_1}\rangle|1_{c_2}\rangle|1_{t}\rangle.
\end{split}
\end{align}

First, the $X_A$ gate acts on $c_2$ to achieve $|1_{c_2}\rangle \stackrel{X_A}{\longleftrightarrow} |2_{c_2}\rangle$ and $|0_{c_2}\rangle \stackrel{X_A}{\longleftrightarrow} |0_{c_2}\rangle$. After the first P-SWAP gate is executed on $c_1$ and $c_2$, $|\psi_0\rangle$ becomes
\begin{align}              \label{eq8}
\begin{split}
|\psi_1\rangle=&\alpha_1|0_{c_1}\rangle|0_{c_2}\rangle|0_{t}\rangle+\alpha_2|0_{c_1}\rangle|0_{c_2}\rangle|1_{t}\rangle
               +\alpha_3|0_{c_1}\rangle|2_{c_2}\rangle|0_{t}\rangle+\alpha_4|0_{c_1}\rangle|2_{c_2}\rangle|1_{t}\rangle\\
               &+\alpha_5|0_{c_1}\rangle|1_{c_2}\rangle|0_{t}\rangle+\alpha_6|0_{c_1}\rangle|1_{c_2}\rangle|1_{t}\rangle
               +\alpha_7|1_{c_1}\rangle|2_{c_2}\rangle|0_{t}\rangle+\alpha_8|1_{c_1}\rangle|2_{c_2}\rangle|1_{t}\rangle.
\end{split}
\end{align}

Second, a CNOT gate is applied to $c_1$ and $t$ (which can be achieved by the circuit in the
dotted rectangle), resulting in
\begin{align}              \label{eq9}
\begin{split}
|\psi_2\rangle=&\alpha_1|0_{c_1}\rangle|0_{c_2}\rangle|0_{t}\rangle+\alpha_2|0_{c_1}\rangle|0_{c_2}\rangle|1_{t}\rangle
               +\alpha_3|0_{c_1}\rangle|2_{c_2}\rangle|0_{t}\rangle+\alpha_4|0_{c_1}\rangle|2_{c_2}\rangle|1_{t}\rangle\\
               &+\alpha_5|0_{c_1}\rangle|1_{c_2}\rangle|0_{t}\rangle+\alpha_6|0_{c_1}\rangle|1_{c_2}\rangle|1_{t}\rangle
               +\alpha_7|1_{c_1}\rangle|2_{c_2}\rangle|1_{t}\rangle+\alpha_8|1_{c_1}\rangle|2_{c_2}\rangle|0_{t}\rangle.
\end{split}
\end{align}

Finally, the P-SWAP and $X_A$ gates are applied again. The two operations induce $|\psi_2\rangle$ as the final state
\begin{align}              \label{eq10}
\begin{split}
|\psi_3\rangle=&\alpha_1|0_{c_1}\rangle|0_{c_2}\rangle|0_{t}\rangle+\alpha_2|0_{c_1}\rangle|0_{c_2}\rangle|1_{t}\rangle
               +\alpha_3|0_{c_1}\rangle|1_{c_2}\rangle|0_{t}\rangle+\alpha_4|0_{c_1}\rangle|1_{c_2}\rangle|1_{t}\rangle\\
               &+\alpha_5|1_{c_1}\rangle|0_{c_2}\rangle|0_{t}\rangle+\alpha_6|1_{c_1}\rangle|0_{c_2}\rangle|1_{t}\rangle
               +\alpha_7|1_{c_1}\rangle|1_{c_2}\rangle|1_{t}\rangle+\alpha_8|1_{c_1}\rangle|1_{c_2}\rangle|0_{t}\rangle.
\end{split}
\end{align}

From Equations (\ref{eq7}-\ref{eq10}), one can see that a three-qubit Toffoli gate can be simulated using three nearest-neighbor qubit-qudit gates and two single-qutrit gates.

\subsubsection{Synthesis of $n$-control-qubit Toffoli gate using qudits} \label{Sec2.2.2}

\begin{figure}  [H]  
\centering
\includegraphics[width=10 cm,angle=0]{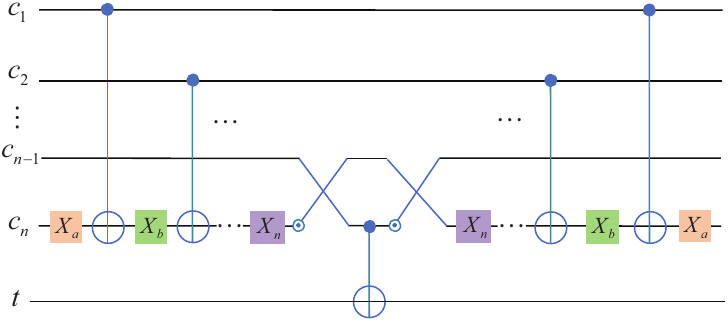}
\caption{Synthesis of an $n$-control-qubit Toffoli gate in a higher-dimensional space. Single-qudit gates $X_a$, $X_b$, $\cdots$, $X_n$ complete operations $|0_{c_n}\rangle\leftrightarrow |2_{c_n}\rangle$, $|1_{c_n}\rangle\leftrightarrow |3_{c_n}\rangle$, $\cdots$, $|0_{c_n}\rangle\leftrightarrow |n_{c_n}\rangle$ when $n$ is even or $|0_{c_n}\rangle\leftrightarrow |2_{c_n}\rangle$, $|1_{c_n}\rangle\leftrightarrow |3_{c_n}\rangle$, $\cdots$, $|1_{c_n}\rangle\leftrightarrow |n_{c_n}\rangle$ when $n$ is odd, respectively. } \label{n-qubit-Toffoli}
\end{figure}

Using a higher-dimensional space, the method can be applied to any multi-qubit Toffoli gate. As shown in Figure \ref{n-qubit-Toffoli}, an $n$-control-qubit Toffoli gate is constructed with ($2n-1$) qubit-qudit and ($2n-2$) single-qudit gates, which flips the target qubit states $|0\rangle$ and $|1\rangle$ if and only if the $n$ control-qubits are all $|1\rangle$. Here, single-qudit gates $X_a$, $X_b$, $\cdots$, $X_n$ create multi-level qudits on $c_n$ and complete transformations $|0_{c_n}\rangle\leftrightarrow |2_{c_n}\rangle$, $|1_{c_n}\rangle\leftrightarrow |3_{c_n}\rangle$, $|0_{c_n}\rangle\leftrightarrow |4_{c_n}\rangle$, $\cdots$, $|0_{c_n}\rangle\leftrightarrow |n_{c_n}\rangle$ when $n$ is even or $|0_{c_n}\rangle\leftrightarrow |2_{c_n}\rangle$, $|1_{c_n}\rangle\leftrightarrow |3_{c_n}\rangle$, $|0_{c_n}\rangle\leftrightarrow |4_{c_n}\rangle$, $\cdots$, $|1_{c_n}\rangle\leftrightarrow |n_{c_n}\rangle$ when $n$ is odd. These single-qudit gates can temporarily expand the two-dimensional space of $c_n$ to an ($n$+1)-dimensional subspace. All CNOT and P-SWAP gates act on computational states $|0\rangle$  and $|1\rangle$. The synthesis requires only $O(n)$  qubit-qudit gates and the low-cost advantage is more evident in our scheme as the number of qubits increases.

\section{Implementation of CNOT and Toffoli gates with linear optics} \label{Sec3}

\subsection{Implementation of a post-selected  P-SWAP gate with linear optics} \label{Sec3.1}

In the previous section, we proposed the simulation of CNOT and Toffoli gates based on P-SWAP gates and auxiliary higher-dimensional spaces. In an optical system, two computational states can be encoded on the polarization DOF of a single photon in the spatial-mode $i$, that is, $|0\rangle \equiv |H\rangle_i$ and $|1\rangle \equiv |V\rangle_i$. Here, $H$ and $V$ represent the horizontal and vertical polarized components, respectively. The higher-dimensional state can be encoded on the $V$-polarized component in a new spatial-mode $i'$, that is, $|2\rangle \equiv |V\rangle_{i'}$. The qutrit operation $X_A$ can be achieved by employing a polarizing beam splitter (PBS), which reflects the $V$-polarized component and transmits the $H$-polarized component, respectively. Before describing the implementation of the CNOT gate, we first detail the step-by-step construction of the P-SWAP gate with linear optical elements.

\begin{figure} 
\begin{center}
\includegraphics[width=8cm,angle=0]{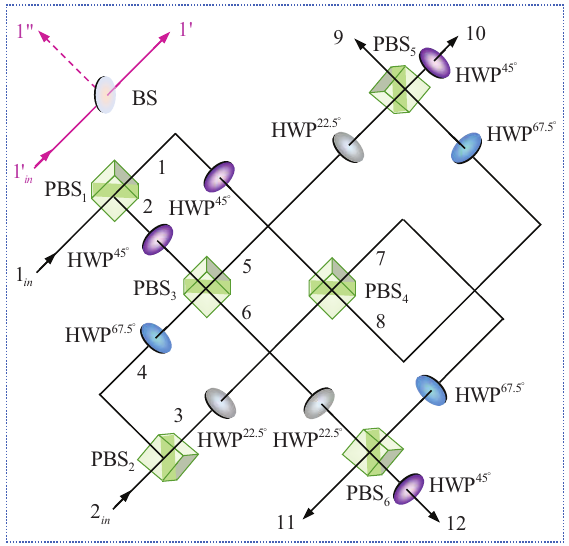}
\caption{Implementation of a linear optical P-SWAP gate. The half-wave plate HWP$^{45^\circ}$ realizes the qubit-flip $\hat{a}^\dag_H\leftrightarrow\hat{a}^\dag_V$. HWP$^{22.5^\circ}$ completes the operations $\hat{a}^\dag_H\leftrightarrow\frac{1}{\sqrt{2}}(\hat{a}^\dag_H+\hat{a}^\dag_V)$ and $\hat{a}^\dag_V\leftrightarrow\frac{1}{\sqrt{2}}(\hat{a}^\dag_H-\hat{a}^\dag_V)$, and HWP$^{67.5^\circ}$ completes $\hat{a}^\dag_H\leftrightarrow\frac{1}{\sqrt{2}}(-\hat{a}^\dag_H+\hat{a}^\dag_V)$  and $\hat{a}^\dag_V\leftrightarrow\frac{1}{\sqrt{2}}(\hat{a}^\dag_H+\hat{a}^\dag_V)$. BS is a balanced beam splitter to realize $\hat{a}^\dag_{V_{{1^{'}{in}}}} \leftrightarrow\frac{1}{\sqrt{2}}(\hat{a}^\dag_{V_{1^{'}}}+\hat{a}^\dag_{V_{1^{''}}})$.} \label{P-SWAP}
\end{center}
\end{figure}

As shown in Figure \ref{P-SWAP}, the injected photon 1 is divided into $H$-polarized component and $V$-polarized component by a PBS. The $H$-polarized component passes into the spatial-mode $1_{in}$, which is encoded on $|H\rangle_{1_{in}}\equiv |0\rangle$  (and $V$-polarized component in the spatial-mode $1_{in}$ is encoded on $|V\rangle_{1_{in}}\equiv |1\rangle$), while the $V$-polarized component is reflected into another spatial-mode $1^{'}_{in}$, which is encoded on $|V\rangle_{1^{'}_{in}} \equiv |2\rangle$. The photon 2  from the spatial-mode $2_{in}$ is encoded on $|H\rangle_{2_{in}} \equiv |0\rangle$ and $|V\rangle_{2_{in}} \equiv |1\rangle$.  A general injected photon state can be considered as
%
\begin{align}              \label{eq11}
\begin{split}
|\varphi_0\rangle=&\big(\alpha_1\hat{a}^\dag_{H_{1_{in}}}\hat{a}^\dag_{H_{2_{in}}}+\alpha_2\hat{a}^\dag_{H_{1_{in}}}\hat{a}^\dag_{V_{2_{in}}}
+\alpha_3\hat{a}^\dag_{V_{1_{in}}}\hat{a}^\dag_{H_{2_{in}}} \\&
+\alpha_4\hat{a}^\dag_{V_{1_{in}}}\hat{a}^\dag_{V_{2_{in}}} +\alpha_5\hat{a}^\dag_{V_{1^{'}{in}}}\hat{a}^\dag_{H_{2_{in}}} +\alpha_6\hat{a}^\dag_{V_{1^{'}{in}}}\hat{a}^\dag_{V_{2_{in}}}\big)|\text{vac}.\rangle.
\end{split}
\end{align}
Here $|\text{vac}.\rangle$ is the state vector of vacuum.

First, PBS$_1$ and PBS$_2$ transmit the $H$-photons into modes 1 and 3 to interact with half-wave plates HWP$^{45^\circ}$ and HWP$^{22.5^\circ}$ and reflect the $V$-photons into modes 2 and 4 to interact with HWP$^{45^\circ}$ and HWP$^{67.5^\circ}$. Here, HWP$^{45^\circ}$ is a half-wave plate set to 45 degrees and achieves the qubit-flip operation $\hat{a}^\dag_H\leftrightarrow\hat{a}^\dag_V$. HWP$^{22.5^\circ}$ completes the transformations
\begin{eqnarray}                  \label{eq12}
\begin{split}
\hat{a}^\dag_H\stackrel{\text{HWP$^{22.5^\circ}$}}{\longleftrightarrow}\frac{1}{\sqrt{2}}(\hat{a}^\dag_H+\hat{a}^\dag_V), \qquad\;\;
\hat{a}^\dag_V\stackrel{\text{HWP$^{22.5^\circ}$}}{\longleftrightarrow}\frac{1}{\sqrt{2}}(\hat{a}^\dag_H-\hat{a}^\dag_V).
\end{split}
\end{eqnarray}
HWP$^{67.5^\circ}$ results in
\begin{eqnarray}                  \label{eq13}
\begin{split}
\hat{a}^\dag_H\stackrel{\text{HWP$^{67.5^\circ}$}}{\longleftrightarrow}\frac{1}{\sqrt{2}}(-\hat{a}^\dag_H+\hat{a}^\dag_V), \qquad\;\;
\hat{a}^\dag_V\stackrel{\text{HWP$^{67.5^\circ}$}}{\longleftrightarrow}\frac{1}{\sqrt{2}}(\hat{a}^\dag_H+\hat{a}^\dag_V).
\end{split}
\end{eqnarray}
The above operations, PBS$_1$ $\rightarrow$ HWP$^{45^\circ}$ (HWP$^{45^\circ}$) and PBS$_2$ $\rightarrow$ HWP$^{22.5^\circ}$ (HWP$^{67.5^\circ}$) cause $|\varphi_0\rangle$ to become
\begin{align}              \label{eq14}
\begin{split}
|\varphi_1\rangle=&\frac{1}{\sqrt{2}}\big[\alpha_1\hat{a}^\dag_{V_1}(\hat{a}^\dag_{H_3}+\hat{a}^\dag_{V_3})+\alpha_2\hat{a}^\dag_{V_1}(\hat{a}^\dag_{H_4}+\hat{a}^\dag_{V_4})
+\alpha_3\hat{a}^\dag_{H_2}(\hat{a}^\dag_{H_3}+\hat{a}^\dag_{V_3})\\& +\alpha_4\hat{a}^\dag_{H_2}(\hat{a}^\dag_{H_4}+\hat{a}^\dag_{V_4})+\alpha_5\hat{a}^\dag_{V_{1^{'}{in}}}(\hat{a}^\dag_{H_3}+\hat{a}^\dag_{V_3}) +\alpha_6\hat{a}^\dag_{V_{1^{'}{in}}}(\hat{a}^\dag_{H_4}+\hat{a}^\dag_{V_4})\big]|\text{vac}.\rangle.
\end{split}
\end{align}

Second, photons in mode $1'_{in}$ are then split into modes $1'$ and $1''$ by a balanced polarization beam splitter (BS), i.e., $\hat{a}^\dag_{V_{{1^{'}{in}}}} \xrightarrow{\text{BS}}(\hat{a}^\dag_{V_{1^{'}}}+\hat{a}^\dag_{V_{1^{''}}})/\sqrt{2}$. Photons emitted from modes 2 and 4 (1 and 3) are split into modes 5 and 6 (7 and 8) by PBS$_3$ (PBS$_4$) and followed  by HWP$^{22.5^\circ}$ (HWP$^{67.5^\circ}$). These elements change $|\varphi_1\rangle$ as
\begin{align}              \label{eq15}
\begin{split}
|\varphi_2\rangle=&\frac{1}{{2\sqrt 2}}\big[
{\alpha _1}(\hat a_{{H_{7}}}^\dag  + \hat a_{{V_{7}}}^\dag )(-\hat a_{{H_7}}^\dag  + \hat a_{{V_{7}}}^\dag + \hat a_{{H_{8}}}^\dag  + \hat a_{{V_{8}}}^\dag)
 + {\alpha _2}(\hat a_{{H_{7}}}^\dag  + \hat a_{{V_{7}}}^\dag )(\hat a_{{H_5}}^\dag +\hat a_{{V_{5}}}^\dag + \hat a_{{H_{6}}}^\dag - \hat a_{{V_{6}}}^\dag)\\&
 + {\alpha _3}(\hat a_{{H_{6}}}^\dag  + \hat a_{{V_{6}}}^\dag )(-\hat a_{{H_7}}^\dag +\hat a_{{V_{7}}}^\dag + \hat a_{{H_{8}}}^\dag  + \hat a_{{V_{8}}}^\dag)
 + {\alpha _4}(\hat a_{{H_{6}}}^\dag  + \hat a_{{V_{6}}}^\dag )(\hat a_{{H_5}}^\dag +\hat a_{{V_{5}}}^\dag + \hat a_{{H_{6}}}^\dag - \hat a_{{V_{6}}}^\dag)\\&
 + {\alpha _5}(\hat a_{{V_{1'}}}^\dag + \hat a_{{V_{1''}}}^\dag)(-\hat a_{{H_7}}^\dag+\hat a_{{V_{7}}}^\dag + \hat a_{{H_{8}}}^\dag  + \hat a_{{V_{8}}}^\dag)
 + {\alpha _6}(\hat a_{{V_{1'}}}^\dag + \hat a_{{V_{1''}}}^\dag)(\hat a_{{H_5}}^\dag +\hat a_{{V_{5}}}^\dag + \hat a_{{H_{6}}}^\dag - \hat a_{{V_{6}}}^\dag)\big]|\text{vac}.\rangle.
\end{split}
\end{align}

Third, PBS$_5$ (PBS$_6$) induces photons into modes 9 and 10 (11 and 12). Photons in modes 10 and 12 will undergo HWP$^{45^\circ}$.  Thus, the state of the system evolves as
\begin{align}              \label{eq16}
\begin{split}
|\varphi_3\rangle=&\frac{1}{{2\sqrt 2}}\big[
{\alpha _1}(\hat a_{{H_{11}}}^\dag  + \hat a_{{H_{12}}}^\dag )(-\hat a_{{H_{11}}}^\dag  + \hat a_{{H_{12}}}^\dag + \hat a_{{H_{9}}}^\dag  + \hat a_{{H_{10}}}^\dag)
 + {\alpha _2}(\hat a_{{H_{11}}}^\dag  + \hat a_{{H_{12}}}^\dag )(\hat a_{V_{10}}^\dag +\hat a_{{V_{9}}}^\dag + \hat a_{{V_{12}}}^\dag - \hat a_{{V_{11}}}^\dag)\\&
 + {\alpha _3}(\hat a_{{V_{12}}}^\dag  + \hat a_{{V_{11}}}^\dag )(-\hat a_{H_{11}}^\dag +\hat a_{{H_{12}}}^\dag + \hat a_{{H_{9}}}^\dag  + \hat a_{{H_{10}}}^\dag)
 + {\alpha _4}(\hat a_{{V_{12}}}^\dag  + \hat a_{{V_{11}}}^\dag )(\hat a_{V_{10}}^\dag +\hat a_{{V_{9}}}^\dag + \hat a_{{V_{12}}}^\dag - \hat a_{{V_{11}}}^\dag)\\&
 + {\alpha _5}(\hat a_{{V_{1'}}}^\dag + \hat a_{{V_{1''}}}^\dag)(-\hat a_{H_{11}}^\dag+\hat a_{{H_{12}}}^\dag + \hat a_{{H_{9}}}^\dag  + \hat a_{{H_{10}}}^\dag)
 + {\alpha _6}(\hat a_{{V_{1'}}}^\dag + \hat a_{{V_{1''}}}^\dag)(\hat a_{V_{10}}^\dag +\hat a_{{V_{9}}}^\dag + \hat a_{{V_{12}}}^\dag - \hat a_{{V_{11}}}^\dag)\big]|\text{vac}.\rangle.
\end{split}
\end{align}
The state $|\varphi_3\rangle$ also has the form
\begin{align}              \label{eq17}
\begin{split}
|\varphi_3\rangle=&|\varphi_4^{1}\rangle+|\varphi_4^{2}\rangle+|\varphi_4^{3}\rangle+|\varphi_4^{4}\rangle
+\frac{1}{{2\sqrt 2}}\big[
{\alpha _1}(\hat a_{{H_{11}}}^\dag  + \hat a_{{H_{12}}}^\dag )(-\hat a_{{H_{11}}}^\dag  + \hat a_{{H_{12}}}^\dag )\\&
+ {\alpha _2}(-\hat a_{{H_{11}}}^\dag\hat a_{{V_{11}}}^\dag + \hat a_{{H_{11}}}^\dag\hat a_{{V_{12}}}^\dag - \hat a_{{V_{11}}}^\dag\hat a_{{H_{12}}}^\dag + \hat a_{{H_{12}}}^\dag\hat a_{{V_{12}}}^\dag) \\&
+ {\alpha _3}(-\hat a_{{H_{11}}}^\dag \hat a_{{V_{11}}}^\dag + \hat a_{{V_{11}}}^\dag \hat a_{{H_{12}}}^\dag - \hat a_{{H_{11}}}^\dag \hat a_{{V_{12}}}^\dag + \hat a_{{H_{12}}}^\dag \hat a_{{V_{12}}}^\dag )\\&
 + {\alpha _4}(\hat a_{{V_{12}}}^\dag  + \hat a_{{V_{11}}}^\dag )(\hat a_{{V_{12}}}^\dag - \hat a_{{V_{11}}}^\dag)
 + {\alpha _5}(\hat a_{{V_{1'}}}^\dag + \hat a_{{V_{1''}}}^\dag)( \hat a_{{H_{9}}}^\dag  + \hat a_{{H_{10}}}^\dag)\\&
 + {\alpha _6}(\hat a_{{V_{1'}}}^\dag + \hat a_{{V_{1''}}}^\dag)(\hat a_{V_{9}}^\dag +\hat a_{{V_{10}}}^\dag)\big]|\text{vac}.\rangle.
\end{split}
\end{align}
Here the four orthogonal states $|\varphi_4^{1}\rangle$, $|\varphi_4^{2}\rangle$, $|\varphi_4^{3}\rangle$, and $|\varphi_4^{4}\rangle$  are given by
\begin{align}              \label{eq18}
\begin{split}
|\varphi_4^{1}\rangle=&\frac{1}{2\sqrt{2}}(\alpha_1\hat a_{H_{9}}^\dag\hat a_{H_{12}}^\dag      +\alpha_2\hat a_{V_{9}}^\dag \hat a_{H_{12}}^\dag
                                          +\alpha_3\hat a_{H_{9}}^\dag\hat a_{V_{12}}^\dag   +\alpha_4\hat a_{V_{9}}^\dag \hat a_{V_{12}}^\dag
                                          +\alpha_5\hat a_{V_{1'}}^\dag\hat a_{H_{12}}^\dag     +\alpha_6\hat a_{V_{1'}}^\dag \hat a_{V_{12}}^\dag)|\text{vac}.\rangle,
\end{split}
\end{align}
\begin{align}              \label{eq19}
\begin{split}
|\varphi_4^{2}\rangle=&\frac{1}{2\sqrt{2}}(\alpha_1\hat a_{H_{10}}^\dag    \hat a_{H_{12}}^\dag      +\alpha_2\hat a_{V_{10}}^\dag \hat a_{H_{12}}^\dag
                                          +\alpha_3\hat a_{H_{10}}^\dag    \hat a_{V_{12}}^\dag    +\alpha_4\hat a_{V_{10}}^\dag \hat a_{V_{12}}^\dag
                                          +\alpha_5\hat a_{V_{1^{''}}}^\dag\hat a_{H_{12}}^\dag      +\alpha_6\hat a_{V_{1^{''}}}^\dag \hat a_{V_{12}}^\dag)|\text{vac}.\rangle,
\end{split}
\end{align}
\begin{align}              \label{eq20}
\begin{split}
|\varphi_4^{3}\rangle=&\frac{1}{2\sqrt{2}}(\alpha_1\hat a_{H_{9}}^\dag \hat a_{H_{11}}^\dag      +\alpha_2\hat a_{V_{9}}^\dag \hat a_{H_{11}}^\dag
                                          +\alpha_3\hat a_{H_{9}}^\dag \hat a_{V_{11}}^\dag    +\alpha_4\hat a_{V_{9}}^\dag \hat a_{V_{11}}^\dag
                                          -\alpha_5\hat a_{V_{1'}}^\dag \hat a_{H_{11}}^\dag     -\alpha_6\hat a_{V_{1'}}^\dag \hat a_{V_{11}}^\dag)|\text{vac}.\rangle,
\end{split}
\end{align}
\begin{align}              \label{eq21}
\begin{split}
|\varphi_4^{4}\rangle=&\frac{1}{2\sqrt{2}}(\alpha_1\hat a_{H_{10}}^\dag \hat a_{H_{11}}^\dag       +\alpha_2 \hat a_{V_{10}}^\dag \hat a_{H_{11}}^\dag
                                          +\alpha_3\hat a_{H_{10}}^\dag \hat a_{V_{11}}^\dag     +\alpha_4 \hat a_{V_{10}}^\dag \hat a_{V_{11}}^\dag
                                          -\alpha_5\hat a_{V_{1''}}^\dag \hat a_{H_{11}}^\dag      -\alpha_6 \hat a_{V_{1''}}^\dag \hat a_{V_{11}}^\dag)|\text{vac}.\rangle.
\end{split}
\end{align}
Based on Equations (\ref{eq18}-\ref{eq21}), a post-selected P-SWAP gate can be operated correctly in the coincidence basis  where the photons act as their own qubits. This means the success of the gate is heralded by a detection of an outing single photon in the desired output ports of the gate (see Table \ref{table1}).

(i) If desired coincidence detections of the output modes are 9, $1'$, and 12 (outputs $1''$, 10, and 11 are discarded),  the state $|\varphi_3\rangle$ will collapse into $|\varphi_4^{1}\rangle$,  and the P-SWAP gate is completed.

(ii) If desired  coincidence detections of the output modes are 10, $1''$, and 12 (outputs $1'$, 9, and 11 are discarded),  the state $|\varphi_3\rangle$ will collapse into
$|\varphi_4^{2}\rangle$,  and the P-SWAP gate is completed.

(iii) If desired coincidence detections of the output modes are 9, $1'$, and 11 (outputs $1''$, 10, and 12 are discarded),  the state $|\varphi_3\rangle$ will collapse into
$|\varphi_4^{3}\rangle$. And then, a phase flip operation, $\hat a^\dag_{V_{1'}} \xrightarrow{\sigma_{z}}  -\hat a^\dag_{V_{1'}}$, should be applied to complete the P-SWAP gate. Such feed-forward operation $\sigma_{z}$ can be easily achieved by setting an HWP$^{0^\circ}$ in output mode $1'$. The feed-forward operations can be determined by post-selection principle, moreover, the spatial-mode-based feed-forward operations have been experimentally demonstrated recently \cite{opti-CNOT1,opti-CNOT4,opti-CNOT5,gao-teleportation}.

(iv) If desired coincidence detections of the output modes are 10, $1''$, and 11 (outputs $1'$, 9, and 12 are discarded),  the state $|\varphi_3\rangle$ will collapse into $|\varphi_4^{4}\rangle$. And then, an HWP$^{0^\circ}$ is set in spatial mode $1''$ to complete the P-SWAP gate.

Putting all the pieces together one can find that the quantum circuit shown in Figure \ref{P-SWAP} completes a linear optical P-SWAP gate in the coincidence basis with a success probability of $4\times1/8=1/2$. The success (or the output modes) of the scheme can be heralded by using the success instances in the post-selection in the applications.

\begin{table*}  
\centering
\caption{Coincident expectation outgoing values for six logic basis inputs. }
\begin{tabular}{ccccccc}
\hline
\multirow{4}{*}{Input}
& $\hat a^\dag_{H_{9}}\hat a^\dag_{H_{12}}$    &\qquad\quad $a^\dag_{V_{9}}a^\dag_{H_{12}}$    &\qquad\quad $a^\dag_{H_{9}}a^\dag_{V_{12}}$     & \qquad\quad $a^\dag_{V_{9}}a^\dag_{V_{12}}$   &\qquad\quad $a^\dag_{V_{1'}}a^\dag_{H_{12}}$   &\qquad\quad $a^\dag_{V_{1'}}a^\dag_{V_{12}}$  \\

\cline{2-7}
 & $a^\dag_{H_{10}}a^\dag_{H_{12}}$   &\qquad\quad $a^\dag_{V_{10}}a^\dag_{H_{12}}$      &\qquad\quad  $a^\dag_{H_{10}}a^\dag_{V_{12}}$     & \qquad\quad $a^\dag_{V_{10}}a^\dag_{V_{12}}$   &\qquad\quad $a^\dag_{V_{1''}}a^\dag_{H_{12}}$   &\qquad\quad $a^\dag_{V_{1''}}a^\dag_{V_{12}}$  \\

\cline{2-7}
& $a^\dag_{H_{9}}a^\dag_{H_{11}}$   &\qquad\quad $a^\dag_{V_{9}}a^\dag_{H_{11}}$      &\qquad\quad $a^\dag_{H_{9}}a^\dag_{V_{11}}$     &\qquad\quad $a^\dag_{V_{9}}a^\dag_{V_{11}}$   &\qquad\quad $-a^\dag_{V_{1'}}a^\dag_{H_{11}}$   &\qquad\quad $-a^\dag_{V_{1'}}a^\dag_{V_{11}}$  \\

\cline{2-7}
& $a^\dag_{H_{10}}a^\dag_{H_{11}}$   &\qquad\quad $a^\dag_{V_{10}}a^\dag_{H_{11}}$      &\qquad\quad $a^\dag_{H_{10}}a^\dag_{V_{11}}$     &\qquad\quad $a^\dag_{V_{10}}a^\dag_{V_{11}}$   &\qquad\quad $-a^\dag_{V_{1''}}a^\dag_{H_{11}}$   &\qquad\quad $-a^\dag_{V_{1''}}a^\dag_{V_{11}}$  \\

\hline

$a^\dag_{H_{1_{in}}}a^\dag_{H_{2_{in}}}$   & 1/8  &\qquad\quad   0  &\qquad\quad 0    &\qquad\quad 0   &\qquad\quad 0    &\qquad\quad 0   \\

$a^\dag_{H_{1_{in}}}a^\dag_{V_{2_{in}}}$    &  0   &\qquad\quad  1/8  &\qquad\quad 0   &\qquad\quad 0   &\qquad\quad 0    &\qquad\quad  0   \\

$a^\dag_{V_{1_{in}}}a^\dag_{H_{2_{in}}}$    & 0    &\qquad\quad   0  &\qquad\quad 1/8 &\qquad\quad 0   &\qquad\quad 0    &\qquad\quad 0  \\

$a^\dag_{V_{1_{in}}}a^\dag_{V_{2_{in}}}$    & 0    &\qquad \quad 0   &\qquad\quad 0  &\qquad\quad 1/8  &\qquad\quad 0    &\qquad\quad  0 \\

$a^\dag_{V_{1'_{in}}}a^\dag_{H_{2_{in}}}$  & 0    &\qquad\quad   0  &\qquad\quad 0    &\qquad\quad 0   &\qquad\quad 1/8  &\qquad\quad  0 \\

$a^\dag_{V_{1'_{in}}}a^\dag_{V_{2_{in}}}$   &0    &\qquad\quad   0  &\qquad\quad 0   &\qquad\quad 0    &\qquad\quad 0    &\qquad\quad 1/8 \\

\hline 
\end{tabular}\label{table1}
\end{table*}


\subsection{Implementation of a post-selected CNOT gate with linear optics} \label{Sec3.2}

\begin{figure}  [H]    %
\centering
\includegraphics[width=10 cm,angle=0]{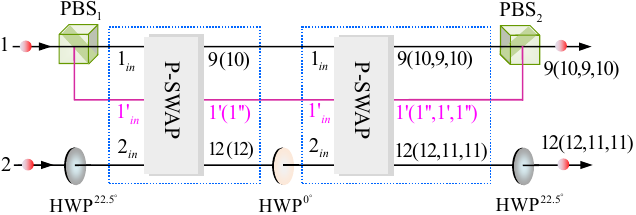}
\caption{Implementation of a linear optical CNOT gate.  The operation in the dotted rectangle is a P-SWAP gate shown in Fig. \ref{P-SWAP}.
Input-output ports pairs mapping $(1_{in}, 1'_{in}, 2_{in})\rightarrow(9, 1', 12)$ and
                                 $(1_{in}, 1'_{in}, 2_{in})\rightarrow (10, 1'', 12)$ are necessary for the leftmost P-SWAP gate.
The outing photons from ports (9, $1'$, 12) and (10, $1''$, 12) of the leftmost P-SWAP gate as inputs will route to the next operations.
Ports pairs mapping $(1_{in}, 1'_{in}, 2_{in})\rightarrow(9, 1', 12)$,
                    $(1_{in}, 1'_{in}, 2_{in})\rightarrow(10, 1'', 12)$,
                    $(1_{in}, 1'_{in}, 2_{in})\rightarrow(9, 1', 11)$, and
                    $(1_{in}, 1'_{in}, 2_{in})\rightarrow(10, 1'', 11)$ are employed for the rightmost P-SWAP gate.
The outing photon pairs emitted from ports pairs $(9, 12)$, $(10, 12)$, $(9, 11)$, and $(10, 11)$ complete the CNOT operation with a success probability of $1/8$. }  \label{optical-CNOT}
\end{figure}

As shown in Figure \ref{optical-CNOT}, a post-selected CNOT gate based on P-SWAP gate can be realized in the coincidence basis with linear optical elements. PBS plays a role in the qutrit $X_A$ to provide an additional spatial mode. The operation in the blue dotted rectangle corresponds to a P-SWAP gate in Figure \ref{P-SWAP}.

First, after the two photons are injected into modes 1 and 2, the input state of the system is given by
\begin{align}
\begin{split}             \label{eq22}
|\chi_0\rangle=&\big(\alpha_1\hat{a}^\dag_{H_{1}}\hat{a}^\dag_{H_{2}}+\alpha_2\hat{a}^\dag_{H_{1}}\hat{a}^\dag_{V_{2}}
+\alpha_3\hat{a}^\dag_{V_{1}}\hat{a}^\dag_{H_{2}}  +\alpha_4\hat{a}^\dag_{V_{1}}\hat{a}^\dag_{V_{2}}\big)|\text{vac}.\rangle.
\end{split}
\end{align}

Second, photons 1 and 2 execute a PBS$_1$ and an HWP$^{22.5^\circ}$, respectively, to pass through the leftmost P-SWAP gate. The output ports of the leftmost P-SWAP gate are modes 9, $1'$, and 12 (or 10, $1''$, and 12), which as an input  will be led to the next HWP$^{0^\circ}$ and the rightmost P-SWAP gate.  PBS$_1$, HWP$^{22.5^\circ}$, and the leftmost P-SWAP gate change $|\chi_0\rangle$ into $|\chi_{9,1',12}\rangle$ or $|\chi_{10,1'',12}\rangle$. Here,
\begin{align}
\begin{split}             \label{eq23}
|\chi_{9,1',12}\rangle_1=&\frac{1}{4}\big[\alpha_1(\hat{a}^\dag_{H_{9}}+\hat{a}^\dag_{V_{9}})\hat{a}^\dag_{H_{12}}
+\alpha_2(\hat{a}^\dag_{H_{9}}-\hat{a}^\dag_{V_{9}})\hat{a}^\dag_{H_{12}} \\&
+\alpha_3\hat{a}^\dag_{V_{1'}}(\hat{a}^\dag_{H_{12}}+\hat{a}^\dag_{V_{12}})
+\alpha_4\hat{a}^\dag_{V_{1'}}(\hat{a}^\dag_{H_{12}}-\hat{a}^\dag_{V_{12}})\big]|\text{vac}.\rangle,
\end{split}
\end{align}
\begin{eqnarray}
\begin{split}             \label{eq24}
\hspace{-1.5em}|\chi_{10,1'',12}\rangle_1= &\frac{1}{4}\big[\alpha_1(\hat{a}^\dag_{H_{10}}+\hat{a}^\dag_{V_{10}})\hat{a}^\dag_{H_{12}}
+\alpha_2(\hat{a}^\dag_{H_{10}}-\hat{a}^\dag_{V_{10}})\hat{a}^\dag_{H_{12}}\\&
+\alpha_3\hat{a}^\dag_{V_{1''}}(\hat{a}^\dag_{H_{12}}+\hat{a}^\dag_{V_{12}})
+\alpha_4\hat{a}^\dag_{V_{1''}}(\hat{a}^\dag_{H_{12}}-\hat{a}^\dag_{V_{12}})\big]|\text{vac}.\rangle.
\end{split}
\end{eqnarray}

Third, HWP$^{0^\circ}$ acts on mode 12 to complete $\hat{a}^\dag_{H_{12}} \rightarrow \hat{a}^\dag_{H_{12}}$ and $\hat{a}^\dag_{V_{12}} \rightarrow -\hat{a}^\dag_{V_{12}}$. The second P-SWAP gate produces eight desired outcomes of the system, that is,
(i) if the desired coincidence detections of the output modes in the second P-SWAP gate are 9, $1'$, and 12,  the state $|\chi_{9,1',12}\rangle_1$ and $|\chi_{10,1'',12}\rangle_1$ both become
\begin{align}
\begin{split}             \label{eq25}
|\chi^+_{9,1',12}\rangle_2=&\frac{1}{8\sqrt{2}}\big[\alpha_1\hat{a}^\dag_{H_{9}}(\hat{a}^\dag_{H_{12}}+\hat{a}^\dag_{V_{12}})
+\alpha_2\hat{a}^\dag_{H_{9}}(\hat{a}^\dag_{H_{12}}-\hat{a}^\dag_{V_{12}})\\&
+\alpha_3\hat{a}^\dag_{V_{1'}}(\hat{a}^\dag_{H_{12}}-\hat{a}^\dag_{V_{12}})
+\alpha_4\hat{a}^\dag_{V_{1'}}(\hat{a}^\dag_{H_{12}}+\hat{a}^\dag_{V_{12}})\big]|\text{vac}.\rangle.
\end{split}
\end{align}
(ii) If the desired coincidence detections of the output modes in the
second P-SWAP gate are 10, $1''$, and 12,  the state $|\chi_{9,1',12}\rangle_1$ and $|\chi_{10,1'',12}\rangle_1$ both become
\begin{eqnarray}
\begin{split}             \label{eq26}
\hspace{-1.5em}|\chi^+_{10,1'',12}\rangle_2=&\frac{1}{8\sqrt{2}}\big[\alpha_1\hat{a}^\dag_{H_{10}}(\hat{a}^\dag_{H_{12}}+\hat{a}^\dag_{V_{12}})
+\alpha_2\hat{a}^\dag_{H_{10}}(\hat{a}^\dag_{H_{12}}-\hat{a}^\dag_{V_{12}})\\&
+\alpha_3\hat{a}^\dag_{V_{1''}}(\hat{a}^\dag_{H_{12}}-\hat{a}^\dag_{V_{12}})
+\alpha_4\hat{a}^\dag_{V_{1''}}(\hat{a}^\dag_{H_{12}}+\hat{a}^\dag_{V_{12}})\big]|\text{vac}.\rangle.
\end{split}
\end{eqnarray}
(iii) If the desired coincidence detections of the output modes in the
second P-SWAP gate are 9, $1''$, and 11, the state $|\chi_{9,1',12}\rangle_1$ and $|\chi_{10,1'',12}\rangle_1$ both become
\begin{align}
\begin{split}             \label{eq27}
|\chi^-_{9,1',11}\rangle_2=&\frac{1}{8\sqrt{2}}\big[\alpha_1\hat{a}^\dag_{H_{9}}(\hat{a}^\dag_{H_{11}}+\hat{a}^\dag_{V_{11}})
+\alpha_2\hat{a}^\dag_{H_{9}}(\hat{a}^\dag_{H_{11}}-\hat{a}^\dag_{V_{11}})\\&
-\alpha_3\hat{a}^\dag_{V_{1'}}(\hat{a}^\dag_{H_{11}}-\hat{a}^\dag_{V_{11}})
-\alpha_4\hat{a}^\dag_{V_{1'}}(\hat{a}^\dag_{H_{11}}+\hat{a}^\dag_{V_{11}})\big]|\text{vac}.\rangle.
\end{split}
\end{align}
(iv) If the desired coincidence detections of the output modes in the
second P-SWAP gate are 10, $1''$, and 11, the state $|\chi_{9,1',12}\rangle_1$ and $|\chi_{10,1'',12}\rangle_1$ both become
\begin{eqnarray}
\begin{split}             \label{eq28}
\hspace{-1.5em}|\chi^-_{10,1'',11}\rangle_2=&\frac{1}{8\sqrt{2}}\big[\alpha_1\hat{a}^\dag_{H_{10}}(\hat{a}^\dag_{H_{11}}+\hat{a}^\dag_{V_{11}})
+\alpha_2\hat{a}^\dag_{H_{10}}(\hat{a}^\dag_{H_{11}}-\hat{a}^\dag_{V_{11}})\\&
-\alpha_3\hat{a}^\dag_{V_{1''}}(\hat{a}^\dag_{H_{11}}-\hat{a}^\dag_{V_{11}})
-\alpha_4\hat{a}^\dag_{V_{1''}}(\hat{a}^\dag_{H_{11}}+\hat{a}^\dag_{V_{11}})\big]|\text{vac}.\rangle.
\end{split}
\end{eqnarray}

Fourth, as shown in Figure \ref{optical-CNOT}, PBS$_2$  leads the photons in modes 9 (i.e., $\hat a^\dag_{H_{9}}|\text{vac}.\rangle$) and $1'$ (i.e., $\hat a^\dag_{V_{1'}}|\text{vac}.\rangle$) into one output mode, and combines the photons in modes 10 (i.e., $\hat a^\dag_{H_{10}}|\text{vac}.\rangle$) and $1''$ (i.e., $\hat a^\dag_{V_{1''}}|\text{vac}.\rangle$) into one output mode.  After  PBS$_2$ and HWP$^{22.5^\circ}$, (i) Equation (\ref{eq25}) evolves into two-fold output state
\begin{align}
\begin{split}             \label{eq29}
|\chi^+_{9,12}\rangle_3=&\frac{1}{8}\big(
 \alpha_1\hat a^\dag_{H_{9}}\hat a^\dag_{H_{12}}
+\alpha_2\hat a^\dag_{H_{9}}\hat a^\dag_{V_{12}}
+\alpha_3\hat a^\dag_{V_{9}}\hat a^\dag_{V_{12}}
+\alpha_4\hat a^\dag_{V_{9}}\hat a^\dag_{H_{12}}\big)|\text{vac}.\rangle.
\end{split}
\end{align}
The CNOT gate is completed.
(ii) Equation (\ref{eq26}) evolves into two-fold output state
\begin{align}
\begin{split}             \label{eq30}
|\chi^+_{10,12}\rangle_3=&\frac{1}{8}\big(
  \alpha_1\hat a^\dag_{H_{10}}\hat a^\dag_{H_{12}}
 +\alpha_2\hat a^\dag_{H_{10}}\hat a^\dag_{V_{12}}
 +\alpha_3\hat a^\dag_{V_{10}}\hat a^\dag_{V_{12}}
 +\alpha_4\hat a^\dag_{V_{10}}\hat a^\dag_{H_{12}}\big)|\text{vac}.\rangle.
\end{split}
\end{align}
The CNOT gate is completed.
(iii) Equation (\ref{eq27}) evolves into two-fold output state
\begin{align}
\begin{split}             \label{eq31}
|\chi^-_{9,11}\rangle_3=&\frac{1}{8}\big(
  \alpha_1\hat a^\dag_{H_{9}}\hat a^\dag_{H_{11}}
 +\alpha_2\hat a^\dag_{H_{9}}\hat a^\dag_{V_{11}}
 -\alpha_3\hat a^\dag_{V_{9}}\hat a^\dag_{V_{11}}
 -\alpha_4\hat a^\dag_{V_{9}}\hat a^\dag_{H_{11}}\big)|\text{vac}.\rangle.
\end{split}
\end{align}
And then an HWP$^{0^\circ}$ is set in the output mode 9 to complete the CNOT gate.
(iv) Equation (\ref{eq28}) evolves into two-fold output state
\begin{align}
\begin{split}             \label{eq32}
|\chi^-_{10,11}\rangle_3=&\frac{1}{8}\big(
 \alpha_1\hat a^\dag_{H_{10}}\hat a^\dag_{H_{11}}
+\alpha_2\hat a^\dag_{H_{10}}\hat a^\dag_{V_{11}}
-\alpha_3\hat a^\dag_{V_{10}}\hat a^\dag_{V_{11}}
-\alpha_4\hat a^\dag_{V_{10}}\hat a^\dag_{H_{11}}\big)|\text{vac}.\rangle.
\end{split}
\end{align}
And then an HWP$^{0^\circ}$ is set in the output mode 10 to complete the CNOT gate.

Based on above orthogonal two-fold states $|\chi^+_{9,12}\rangle_3$, $|\chi^+_{10,12}\rangle_3$, $|\chi^-_{9,11}\rangle_3$, and $|\chi^-_{10,11}\rangle_3$, one can find that after the feed-forward operations are only applied to the rightmost P-SWAP gate, an optical post-selected CNOT gate can be operated correctly with a success probability of $8\times1/64=1/8$.   Remarkably, additional entangled photon pairs or single photons are necessary for previous schemes \cite{opti-CNOT1,opti-CNOT3,opti-CNOT4,gao-teleportation,opti-CNOT2,opti-CNOT5}, but are not required for our CNOT gate. In addition, the success probability of the gate is improved on the results without an auxiliary photon \cite{CNOT1/9-5,CNOT1/9-0,CNOT1/9-1,CNOT1/9-2,CNOT1/9-3}. We also note that  other method for implementing linear optical CNOT gate with additional DOFs has been demonstrated experimentally \cite{ZhouCNOT}.

\subsection{Implementation of  a post-selected  Toffoli gate with linear optics} \label{Sec3.3}

\begin{figure}  [H]    %
\centering
\includegraphics[width=9 cm,angle=0]{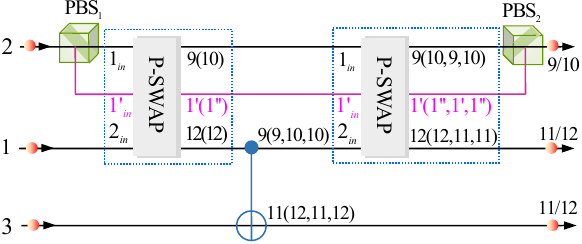}
\caption{Optical implementation of a three-photon Toffoli gate.}  \label{optical-Toffoli}
\end{figure}

We propose the implementation of a Toffoli gate based on the designed P-SWAP and CNOT gates. As shown in Figure \ref{optical-Toffoli},  three photons are injected into modes 1, 2, and 3, simultaneously, the initial state is given by
\begin{align}
\begin{split}             \label{eq33}
|\Xi_{0}\rangle=&\big(
 \alpha_1\hat{a}^\dag_{H_{1}}\hat{a}^\dag_{H_{2}}\hat{a}^\dag_{H_{3}}
+\alpha_2\hat{a}^\dag_{H_{1}}\hat{a}^\dag_{H_{2}}\hat{a}^\dag_{V_{3}}
+\alpha_3\hat{a}^\dag_{H_{1}}\hat{a}^\dag_{V_{2}}\hat{a}^\dag_{H_{3}}
+\alpha_4\hat{a}^\dag_{H_{1}}\hat{a}^\dag_{V_{2}}\hat{a}^\dag_{V_{3}}\\&
+\alpha_5\hat{a}^\dag_{V_{1}}\hat{a}^\dag_{H_{2}}\hat{a}^\dag_{H_{3}}
+\alpha_6\hat{a}^\dag_{V_{1}}\hat{a}^\dag_{H_{2}}\hat{a}^\dag_{V_{3}}
+\alpha_7\hat{a}^\dag_{V_{1}}\hat{a}^\dag_{V_{2}}\hat{a}^\dag_{H_{3}}
+\alpha_8\hat{a}^\dag_{V_{1}}\hat{a}^\dag_{V_{2}}\hat{a}^\dag_{V_{3}}\big)|\text{vac}.\rangle.
\end{split}
\end{align}

First, after photons go through the PBS$_1$ and the leftmost P-SWAP gate, if the output modes of the leftmost P-SWAP gate are 9, $1'$, and 12, (10, $1''$, and 12 ) and we can obtain two desired states,
\begin{align}
\begin{split}             \label{eq34}
|\Xi_{1a}\rangle=&\frac{1}{2\sqrt{2}}\big(\alpha_1\hat{a}^\dag_{H_{12}}\hat{a}^\dag_{H_{9}}\hat{a}^\dag_{H_{3}}
+\alpha_2\hat{a}^\dag_{H_{12}}\hat{a}^\dag_{H_{9}}\hat{a}^\dag_{V_{3}}
+\alpha_3\hat{a}^\dag_{H_{12}}\hat{a}^\dag_{V_{1'}}\hat{a}^\dag_{H_{3}}
+\alpha_4\hat{a}^\dag_{H_{12}}\hat{a}^\dag_{V_{1'}}\hat{a}^\dag_{V_{3}}\\&
+\alpha_5\hat{a}^\dag_{H_{12}}\hat{a}^\dag_{V_{9}}\hat{a}^\dag_{H_{3}}
+\alpha_6\hat{a}^\dag_{H_{12}}\hat{a}^\dag_{V_{9}}\hat{a}^\dag_{V_{3}}
+\alpha_7\hat{a}^\dag_{V_{12}}\hat{a}^\dag_{V_{1'}}\hat{a}^\dag_{H_{3}}
+\alpha_8\hat{a}^\dag_{V_{12}}\hat{a}^\dag_{V_{1'}}\hat{a}^\dag_{V_{3}}\big)|\text{vac}.\rangle,
\end{split}
\end{align}
\begin{align}
\begin{split}             \label{eq35}
|\Xi_{1b}\rangle=&\frac{1}{2\sqrt{2}}\big(\alpha_1\hat{a}^\dag_{H_{12}}\hat{a}^\dag_{H_{10}}\hat{a}^\dag_{H_{3}}
+\alpha_2\hat{a}^\dag_{H_{12}}\hat{a}^\dag_{H_{10}}\hat{a}^\dag_{V_{3}}
+\alpha_3\hat{a}^\dag_{H_{12}}\hat{a}^\dag_{V_{1''}}\hat{a}^\dag_{H_{3}}
+\alpha_4\hat{a}^\dag_{H_{12}}\hat{a}^\dag_{V_{1''}}\hat{a}^\dag_{V_{3}}\\&
+\alpha_5\hat{a}^\dag_{H_{12}}\hat{a}^\dag_{V_{10}}\hat{a}^\dag_{H_{3}}
+\alpha_6\hat{a}^\dag_{H_{12}}\hat{a}^\dag_{V_{10}}\hat{a}^\dag_{V_{3}}
+\alpha_7\hat{a}^\dag_{V_{12}}\hat{a}^\dag_{V_{1''}}\hat{a}^\dag_{H_{3}}
+\alpha_8\hat{a}^\dag_{V_{12}}\hat{a}^\dag_{V_{1''}}\hat{a}^\dag_{V_{3}}\big)|\text{vac}.\rangle.
\end{split}
\end{align}

Second, the states described by Equation (\ref{eq34}) and Equation (\ref{eq35}) are employed as the initial states for the next CNOT gate acting on photon 1 and photon 3. If the outing
photons emitted from path pairs $(9, 11)$,  or $(9, 12)$, or $(10, 11)$, or $(10, 12)$,  which can yield 16 desired states $|\Xi_{i,9,k}\rangle_1$ (two-fold) and $|\Xi_{i,10,k}\rangle_1$ (two-fold). Here, $|\Xi_{i,9,k}\rangle_1$ and $|\Xi_{i,10,k}\rangle_1$ with $i\in\{9, 10\}$ and $k\in\{11, 12\}$ are described by
\begin{align}
\begin{split}             \label{eq36}
|\Xi_{i,9,k}\rangle_1=&\frac{1}{16\sqrt{2}}\big(
 \alpha_1\hat{a}^\dag_{H_{i}}\hat{a}^\dag_{H_{9}}\hat{a}^\dag_{H_{k}}
+\alpha_2\hat{a}^\dag_{H_{i}}\hat{a}^\dag_{H_{9}}\hat{a}^\dag_{V_{k}}
+\alpha_3\hat{a}^\dag_{H_{i}}\hat{a}^\dag_{V_{1'}}\hat{a}^\dag_{H_{k}}
+\alpha_4\hat{a}^\dag_{H_{i}}\hat{a}^\dag_{V_{1'}}\hat{a}^\dag_{V_{k}}\\&
+\alpha_5\hat{a}^\dag_{H_{i}}\hat{a}^\dag_{V_{9}}\hat{a}^\dag_{H_{k}}
+\alpha_6\hat{a}^\dag_{H_{i}}\hat{a}^\dag_{V_{9}}\hat{a}^\dag_{V_{k}}
+\alpha_7\hat{a}^\dag_{V_{i}}\hat{a}^\dag_{V_{1'}}\hat{a}^\dag_{V_{k}}
+\alpha_8\hat{a}^\dag_{V_{i}}\hat{a}^\dag_{V_{1'}}\hat{a}^\dag_{H_{k}}\big)|\text{vac}.\rangle,
\end{split}
\end{align}
\begin{align}
\begin{split}             \label{eq37}
|\Xi_{i,10,k}\rangle_1=&\frac{1}{16\sqrt{2}}\big(
 \alpha_1\hat{a}^\dag_{H_{i}}\hat{a}^\dag_{H_{10}}\hat{a}^\dag_{H_{k}}
+\alpha_2\hat{a}^\dag_{H_{i}}\hat{a}^\dag_{H_{10}}\hat{a}^\dag_{V_{k}}
+\alpha_3\hat{a}^\dag_{H_{i}}\hat{a}^\dag_{V_{1''}}\hat{a}^\dag_{H_{k}}
+\alpha_4\hat{a}^\dag_{H_{i}}\hat{a}^\dag_{V_{1''}}\hat{a}^\dag_{V_{k}}\\&
+\alpha_5\hat{a}^\dag_{H_{i}}\hat{a}^\dag_{V_{10}}\hat{a}^\dag_{H_{k}}
+\alpha_6\hat{a}^\dag_{H_{i}}\hat{a}^\dag_{V_{10}}\hat{a}^\dag_{V_{k}}
+\alpha_7\hat{a}^\dag_{V_{i}}\hat{a}^\dag_{V_{1''}}\hat{a}^\dag_{V_{k}}
+\alpha_8\hat{a}^\dag_{V_{i}}\hat{a}^\dag_{V_{1''}}\hat{a}^\dag_{H_{k}}\big)|\text{vac}.\rangle.
\end{split}
\end{align}

Third, above 16 states  are introduced as the initial states for the rightmost P-SWAP gate acting on photon 1 and photon 2.  If coincidence detection  mode pairs are (9, $1'$, and 12),  or (9, $1'$, and 11),  or (10, $1''$, and 12),  or (10, $1''$, and 11), we can obtain 64 desired states  $|\Xi^+_{12,9,k}\rangle_2$, $|\Xi^+_{12,10,k}\rangle_2$, $|\Xi^-_{11,9,k}\rangle_2$ and $|\Xi^-_{11,10,k}\rangle_2$. Here, eight-fold states $|\Xi^+_{12,9,k}\rangle_2$, $|\Xi^+_{12,10,k}\rangle_2$, $|\Xi^-_{11,9,k}\rangle_2$, and $|\Xi^-_{11,10,k}\rangle_2$  with $k\in\{11, 12\}$ are described by
\begin{align}
\begin{split}             \label{eq38}
|\Xi^+_{12,9,k}\rangle_2=&\frac{1}{64}\big(
  \alpha_1\hat{a}^\dag_{H_{12}}\hat{a}^\dag_{H_{9}} \hat{a}^\dag_{H_{k}}
 +\alpha_2\hat{a}^\dag_{H_{12}}\hat{a}^\dag_{H_{9}} \hat{a}^\dag_{V_{k}}
 +\alpha_3\hat{a}^\dag_{H_{12}}\hat{a}^\dag_{V_{1'}}\hat{a}^\dag_{H_{k}}
 +\alpha_4\hat{a}^\dag_{H_{12}}\hat{a}^\dag_{V_{1'}}\hat{a}^\dag_{V_{k}}\\&
 +\alpha_5\hat{a}^\dag_{V_{12}}\hat{a}^\dag_{H_{9}} \hat{a}^\dag_{H_{k}}
 +\alpha_6\hat{a}^\dag_{V_{12}}\hat{a}^\dag_{H_{9}} \hat{a}^\dag_{V_{k}}
 +\alpha_7\hat{a}^\dag_{V_{12}}\hat{a}^\dag_{V_{1'}}\hat{a}^\dag_{V_{k}}
 +\alpha_8\hat{a}^\dag_{V_{12}}\hat{a}^\dag_{V_{1'}}\hat{a}^\dag_{H_{k}}\big)|\text{vac}.\rangle,
\end{split}
\end{align}
\begin{align}
\begin{split}             \label{eq39}
|\Xi^+_{12,10,k}\rangle_2=&\frac{1}{64}\big(
  \alpha_1\hat{a}^\dag_{H_{12}}\hat{a}^\dag_{H_{10}}\hat{a}^\dag_{H_{k}}
 +\alpha_2\hat{a}^\dag_{H_{12}}\hat{a}^\dag_{H_{10}}\hat{a}^\dag_{V_{k}}
+\alpha_3\hat{a}^\dag_{H_{12}}\hat{a}^\dag_{V_{1''}}\hat{a}^\dag_{H_{k}}
 +\alpha_4\hat{a}^\dag_{H_{12}}\hat{a}^\dag_{V_{1''}}\hat{a}^\dag_{V_{k}}\\
&+\alpha_5\hat{a}^\dag_{V_{12}}\hat{a}^\dag_{H_{10}}\hat{a}^\dag_{H_{k}}
 +\alpha_6\hat{a}^\dag_{V_{12}}\hat{a}^\dag_{H_{10}}\hat{a}^\dag_{V_{k}}
+\alpha_7\hat{a}^\dag_{V_{12}}\hat{a}^\dag_{V_{1''}}\hat{a}^\dag_{V_{k}}
 +\alpha_8\hat{a}^\dag_{V_{12}}\hat{a}^\dag_{V_{1''}}\hat{a}^\dag_{H_{k}}\big)|\text{vac}.\rangle,
\end{split}
\end{align}
\begin{align}
\begin{split}             \label{eq40}
|\Xi^-_{11,9,k}\rangle_2=&\frac{1}{64}\big(
  \alpha_1\hat{a}^\dag_{H_{11}}\hat{a}^\dag_{H_{9}}\hat{a}^\dag_{H_{k}}
 +\alpha_2\hat{a}^\dag_{H_{11}}\hat{a}^\dag_{H_{9}}\hat{a}^\dag_{V_{k}}
-\alpha_3\hat{a}^\dag_{H_{11}}\hat{a}^\dag_{V_{1'}}\hat{a}^\dag_{H_{k}}
 -\alpha_4\hat{a}^\dag_{H_{11}}\hat{a}^\dag_{V_{1'}}\hat{a}^\dag_{V_{k}}\\
&+\alpha_5\hat{a}^\dag_{V_{11}}\hat{a}^\dag_{H_{9}}\hat{a}^\dag_{H_{k}}
 +\alpha_6\hat{a}^\dag_{V_{11}}\hat{a}^\dag_{H_{9}}\hat{a}^\dag_{V_{k}}
 -\alpha_7\hat{a}^\dag_{V_{11}}\hat{a}^\dag_{V_{1'}}\hat{a}^\dag_{H_{k}}
 -\alpha_8\hat{a}^\dag_{V_{11}}\hat{a}^\dag_{V_{1'}}\hat{a}^\dag_{V_{k}}\big)|\text{vac}.\rangle,
\end{split}
\end{align}
\begin{align}
\begin{split}             \label{eq41}
|\Xi^-_{11,10,k}\rangle_2=&\frac{1}{64}\big(
   \alpha_1\hat{a}^\dag_{H_{11}}\hat{a}^\dag_{H_{10}}\hat{a}^\dag_{H_{k}}
  +\alpha_2\hat{a}^\dag_{H_{11}}\hat{a}^\dag_{H_{10}}\hat{a}^\dag_{V_{k}}
 -\alpha_3\hat{a}^\dag_{H_{11}}\hat{a}^\dag_{V_{1''}}\hat{a}^\dag_{H_{k}}
  -\alpha_4\hat{a}^\dag_{H_{11}}\hat{a}^\dag_{V_{1''}}\hat{a}^\dag_{V_{k}}\\
 &+\alpha_5\hat{a}^\dag_{V_{11}}\hat{a}^\dag_{H_{10}}\hat{a}^\dag_{H_{k}}
  +\alpha_6\hat{a}^\dag_{V_{11}}\hat{a}^\dag_{H_{10}}\hat{a}^\dag_{V_{k}}
 -\alpha_7\hat{a}^\dag_{V_{11}}\hat{a}^\dag_{V_{1''}}\hat{a}^\dag_{V_{k}}
  -\alpha_8\hat{a}^\dag_{V_{11}}\hat{a}^\dag_{V_{1''}}\hat{a}^\dag_{H_{k}}\big)|\text{vac}.\rangle.
\end{split}
\end{align}

Finally, as shown in Figure \ref{optical-Toffoli}, the photons emitted from modes 9 (i.e., $\hat{a}^\dag_{H_{9}}|\text{vac}.\rangle$) and $1'$ (i.e., $\hat{a}^\dag_{V_{1'}}|\text{vac}.\rangle$) are combined into the same output mode by PBS$_2$. The photons emitted from modes 10 (i.e., $\hat{a}^\dag_{H_{10}}|\text{vac}.\rangle$) and $1''$ ($\hat{a}^\dag_{V_{1''}}|\text{vac}.\rangle$) are also leaded to the same output mode by PBS$_2$.
 Therefore, after PBS$_2$, (i) Equation (\ref{eq38}) evolves into eight-fold output state
\begin{align}
\begin{split}             \label{eq42}
|\Xi^+_{12,9,k}\rangle_3=&\frac{1}{64}\big(
  \alpha_1\hat{a}^\dag_{H_{12}}\hat{a}^\dag_{H_{9}} \hat{a}^\dag_{H_{k}}
 +\alpha_2\hat{a}^\dag_{H_{12}}\hat{a}^\dag_{H_{9}} \hat{a}^\dag_{V_{k}}
 +\alpha_3\hat{a}^\dag_{H_{12}}\hat{a}^\dag_{V_{9}}\hat{a}^\dag_{H_{k}}
 +\alpha_4\hat{a}^\dag_{H_{12}}\hat{a}^\dag_{V_{9}}\hat{a}^\dag_{V_{k}}\\&
 +\alpha_5\hat{a}^\dag_{V_{12}}\hat{a}^\dag_{H_{9}} \hat{a}^\dag_{H_{k}}
 +\alpha_6\hat{a}^\dag_{V_{12}}\hat{a}^\dag_{H_{9}} \hat{a}^\dag_{V_{k}}
 +\alpha_7\hat{a}^\dag_{V_{12}}\hat{a}^\dag_{V_{9}}\hat{a}^\dag_{V_{k}}
 +\alpha_8\hat{a}^\dag_{V_{12}}\hat{a}^\dag_{V_{9}}\hat{a}^\dag_{H_{k}}\big)|\text{vac}.\rangle.
\end{split}
\end{align}
The three-photon Toffoli gate is completed.
(ii)  Equation (\ref{eq39}) evolves into eight-fold output state
\begin{align}
\begin{split}             \label{eq43}
|\Xi^+_{12,10,k}\rangle_3=&\frac{1}{64}\big(
  \alpha_1\hat{a}^\dag_{H_{12}}\hat{a}^\dag_{H_{10}}\hat{a}^\dag_{H_{k}}
 +\alpha_2\hat{a}^\dag_{H_{12}}\hat{a}^\dag_{H_{10}}\hat{a}^\dag_{V_{k}}
+\alpha_3\hat{a}^\dag_{H_{12}}\hat{a}^\dag_{V_{10}}\hat{a}^\dag_{H_{k}}
 +\alpha_4\hat{a}^\dag_{H_{12}}\hat{a}^\dag_{V_{10}}\hat{a}^\dag_{V_{k}}\\
&+\alpha_5\hat{a}^\dag_{V_{12}}\hat{a}^\dag_{H_{10}}\hat{a}^\dag_{H_{k}}
 +\alpha_6\hat{a}^\dag_{V_{12}}\hat{a}^\dag_{H_{10}}\hat{a}^\dag_{V_{k}}
 +\alpha_7\hat{a}^\dag_{V_{12}}\hat{a}^\dag_{V_{10}}\hat{a}^\dag_{V_{k}}
 +\alpha_8\hat{a}^\dag_{V_{12}}\hat{a}^\dag_{V_{10}}\hat{a}^\dag_{H_{k}}\big)|\text{vac}.\rangle.
\end{split}
\end{align}
The three-photon Toffoli gate is also completed.
(iii)  Equation (\ref{eq40}) evolves into eight-fold output state
\begin{align}
\begin{split}             \label{eq44}
|\Xi^-_{11,9,k}\rangle_3=&\frac{1}{64}\big(
  \alpha_1\hat{a}^\dag_{H_{11}}\hat{a}^\dag_{H_{9}}\hat{a}^\dag_{H_{k}}
 +\alpha_2\hat{a}^\dag_{H_{11}}\hat{a}^\dag_{H_{9}}\hat{a}^\dag_{V_{k}}
-\alpha_3\hat{a}^\dag_{H_{11}}\hat{a}^\dag_{V_{9}}\hat{a}^\dag_{H_{k}}
 -\alpha_4\hat{a}^\dag_{H_{11}}\hat{a}^\dag_{V_{9}}\hat{a}^\dag_{V_{k}}\\
&+\alpha_5\hat{a}^\dag_{V_{11}}\hat{a}^\dag_{H_{9}}\hat{a}^\dag_{H_{k}}
 +\alpha_6\hat{a}^\dag_{V_{11}}\hat{a}^\dag_{H_{9}}\hat{a}^\dag_{V_{k}}
-\alpha_7\hat{a}^\dag_{V_{11}}\hat{a}^\dag_{V_{9}}\hat{a}^\dag_{H_{k}}
 -\alpha_8\hat{a}^\dag_{V_{11}}\hat{a}^\dag_{V_{9}}\hat{a}^\dag_{V_{k}}\big)|\text{vac}.\rangle.
\end{split}
\end{align}
And then an HWP$^{0^\circ}$  is set in the output mode 9 to complete the three-photon Toffoli gate.
(iv)  Equation (\ref{eq41}) evolves into eight-fold output state
\begin{align}
\begin{split}             \label{eq45}
|\Xi^-_{11,10,k}\rangle_3=&\frac{1}{64}\big(
   \alpha_1\hat{a}^\dag_{H_{11}}\hat{a}^\dag_{H_{10}}\hat{a}^\dag_{H_{k}}
  +\alpha_2\hat{a}^\dag_{H_{11}}\hat{a}^\dag_{H_{10}}\hat{a}^\dag_{V_{k}}
 -\alpha_3\hat{a}^\dag_{H_{11}}\hat{a}^\dag_{V_{10}}\hat{a}^\dag_{H_{k}}
  -\alpha_4\hat{a}^\dag_{H_{11}}\hat{a}^\dag_{V_{10}}\hat{a}^\dag_{V_{k}}\\
 &+\alpha_5\hat{a}^\dag_{V_{11}}\hat{a}^\dag_{H_{10}}\hat{a}^\dag_{H_{k}}
  +\alpha_6\hat{a}^\dag_{V_{11}}\hat{a}^\dag_{H_{10}}\hat{a}^\dag_{V_{k}}
 -\alpha_7\hat{a}^\dag_{V_{11}}\hat{a}^\dag_{V_{10}}\hat{a}^\dag_{V_{k}}
  -\alpha_8\hat{a}^\dag_{V_{11}}\hat{a}^\dag_{V_{10}}\hat{a}^\dag_{H_{k}}\big)|\text{vac}.\rangle.
\end{split}
\end{align}
And then an HWP$^{0^\circ}$ is set in the output mode 10 to complete the three-photon Toffoli gate.

\begin{figure}  [H]   
\centering
\includegraphics[width=11cm,angle=0]{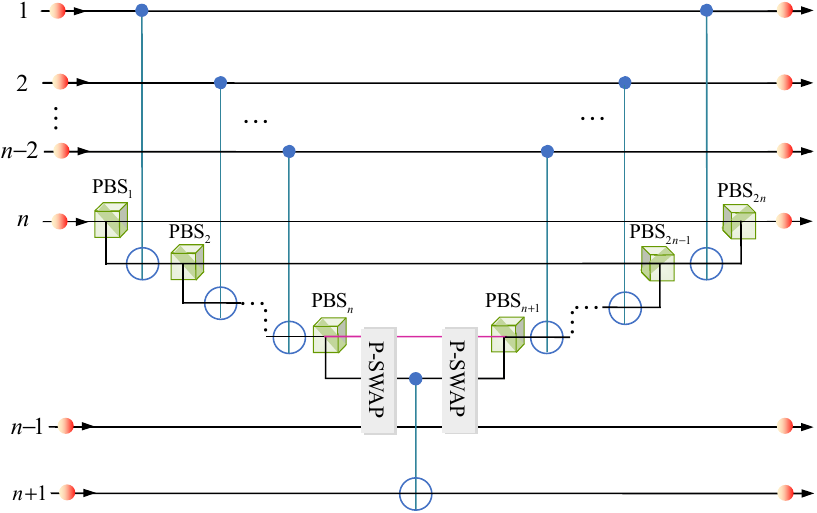}
\caption{Implementation of an ($n+1$)-photon Toffoli gate.}  \label{n-optical-Toffoli}
\end{figure}

Based on above orthogonal eight-fold states described by Equations (\ref{eq42}-\ref{eq45}), one can find that our proposal can be achieved in the coincidence basis with  a higher success probability ($64\times1/64^2=1/64$) than the simplified CNOT-based one (1/72) \cite{T-PRA,T-NatPhy} and the one without a decomposition-based approach (1/133)  \cite{LO-Toffoli}. In addition, optical single-qudit operation ensembles $X_a$, $X_b$, $\cdots$, $X_n$ can be achieved by employing a sequence of PBSs, and the linear optical $n$-control-photon Toffoli gate can be implemented in principle (see Figure \ref{n-optical-Toffoli}).


\section{Discussion and Conclusion}

The optimal cost of a Toffoli gate is six CNOT gates using the standard decomposition-based approach in qubit system \cite{six-Toffoli}. The theoretical lower bound of a Toffoli gate is five two-qubit gates in qubit system \cite{Five1}. Ralph \emph{et al}. \cite{T-PRA} first reduced the cost of a Toffoli gate to three qubit-qudit CNOT gates by introducing a qutrit.  Using the same idea as the works in Refs. \cite{T-PRA,T-NatPhy},  we designed an alternative the quantum circuit to implement the Toffoli gate with a higher success probability based on the P-SWAP gates, which required the same number of qubit-qudit gates as the protocols in Refs. \cite{T-PRA,T-NatPhy}.  The required qubit-qudit entangled  gates are all nearest neighbors in our construction of the three-qubit Toffoli gate. Note that the nearest-neighbor quantum gate where each qubit interacts only with its nearest neighbors requires less resource overhead than the long-range one. For example, a long-range CNOT gate acting on the first qubit and the third qubit is constructed by four nearest-neighbor CNOT gates \cite{long-range}. In addition, ($2n-1$) qubit-qudit gates  and ($2n-2$) single-qudit gates can simulate an $n$-control-qubit Toffoli gate in higher-dimensional spaces.

Linear optics has inherent probability characteristics for the implementation of controlled quantum gates. With the help of an additional entangled photon pair \cite{opti-CNOT3,opti-CNOT4} or a single photon  \cite{opti-CNOT2}, optical CNOT gate with a success probability of 1/4  or 1/8 can be realistically implemented. Without auxiliary photons, CNOT gate with a success probability of 1/9 has been experimentally demonstrated in linear optics \cite{CNOT1/9-0,CNOT1/9-1,CNOT1/9-2,CNOT1/9-3}. Remarkably, the success probability of our P-SWAP-based CNOT gate is enhanced to 1/8 without additional photons. Moreover, the success probability of our P-SWAP-based Toffoli gate (1/64) is higher than the CNOT-based protocols (1/72) \cite{T-PRA,T-NatPhy} and it is also higher than the no-decomposition-based one (1/133) \cite{LO-Toffoli}.

The multi-level system is essential to realize our schemes. In optical system, we can encode polarization DOF of photons as two computational qubits and spatial-mode DOF as the qudit (extra level).
We can also encode these levels on orbital angular momentum of photons. Besides, diamond nitrogen-vacancy defect center \cite{NV1,NV2} and superconducting system \cite{superconducting1,superconducting2}  can also provide available multiple levels to implement the universal quantum gates due to their long coherence time and flexible manipulation.


In summary, by introducing higher-dimensional spaces, we proposed simplified CNOT and Toffoli gates.
A three-qubit Toffoli gate can be simulated with two P-SWAP, one CNOT, and two single-qutrit gates.  $(2n-1)$ qubit-qudit gates  and $(2n-2)$ single-qudit gates are sufficient for constructing an $n$-control-qubit Toffoli gate.
Following the simplified synthesis, as a feasible example, linear optics architectures for implementing CNOT and Toffoli gates were designed with a higher success probability.

\bigskip



\section*{ACKNOWLEDGMENTS}
This work is supported by the National Natural Science Foundation of China under Grant No. 11604012, the Fundamental Research Funds for the Central Universities under Grants No. FRF-TP-19-011A3 and No. 230201506500024, and a grant from the China Scholarship Council.  L.-C.K. is supported by the Ministry of Education and the National Research Foundation Singapore.

%



\end{document}